\documentclass[aps,prl,amsmath,amssymb,superscriptaddress,showpacs,showkeys,twocolumn]{revtex4}
\usepackage{textcomp}
\usepackage{graphicx}
\usepackage{dcolumn}
\usepackage{bm}
\usepackage{siunitx}
\usepackage[utf8]{inputenc}
\usepackage[english]{babel}
\usepackage{natbib}
\usepackage{mathrsfs}
\usepackage{enumerate, mathtools, braket} 
\usepackage[a4paper]{geometry}
\usepackage{hyperref}

\begin{document}


\title{Electromagnetically Induced Transparency of On-demand Single Photons in a Hybrid Quantum Network} 

\author{Lucas~Schweickert}
\altaffiliation{L.~Schweickert and K.~D.~J\"ons contributed equally to this work.}
\author{Klaus~D.~J\"ons}
\email[Corresponding author: ]{klausj@kth.se}
\altaffiliation{L.~Schweickert and K.~D.~J\"ons contributed equally to this work.}
\affiliation{Department of Applied Physics, Royal Institute of Technology, Albanova University Centre, Roslagstullsbacken 21, 106 91 Stockholm, Sweden}%
\author{Mehdi~Namazi}
\affiliation{Department of Physics \& Astronomy, Stony Brook University, Stony Brook, NY 11794-3800, USA}
\author{Guodong~Cui}
\affiliation{Department of Physics \& Astronomy, Stony Brook University, Stony Brook, NY 11794-3800, USA}
\author{Thomas~Lettner}
\author{Katharina~D.~Zeuner}
\author{Lara~Scavuzzo~Monta\~{n}a}
\affiliation{Department of Applied Physics, Royal Institute of Technology, Albanova University Centre, Roslagstullsbacken 21, 106 91 Stockholm, Sweden}%
\author{Saimon~Filipe~Covre~da~Silva} 
\author{Marcus~Reindl}
\author{Huiying~Huang}
\affiliation{Institute of Semiconductor and Solid State Physics, Johannes Kepler University Linz, 4040, Austria}
\author{Rinaldo~Trotta}
\affiliation{Institute of Semiconductor and Solid State Physics, Johannes Kepler University Linz, 4040, Austria}
\affiliation{Dipartimento di Fisica, Sapienza Universit\`a di Roma, Piazzale A. Moro 1, I-00185 Roma, Italy}
\author{Armando~Rastelli}
\affiliation{Institute of Semiconductor and Solid State Physics, Johannes Kepler University Linz, 4040, Austria}
\author{Val~Zwiller}
\affiliation{Department of Applied Physics, Royal Institute of Technology, Albanova University Centre, Roslagstullsbacken 21, 106 91 Stockholm, Sweden}%
\affiliation{Kavli Institute of Nanoscience, Delft University of Technology, Lorentzweg 1, 2628CJ Delft, The Netherlands}
\author{Eden~Figueroa}
\email[Corresponding author: ]{eden.figueroa@stonybrook.edu}
\affiliation{Department of Physics \& Astronomy, Stony Brook University, Stony Brook, NY 11794-3800, USA}

\date{\today}

\begin{abstract}
Long range quantum communication and quantum information processing require the development of light-matter interfaces for distributed quantum networks. Even though photons are ideal candidates for network links to transfer quantum information, the system of choice for the realization of quantum nodes has not been identified yet. Ideally, one strives for a hybrid network architecture, which will consist of different quantum systems, combining the strengths of each system. However, interfacing different quantum systems via photonic channels remains a major challenge because a detailed understanding of the underlying light-matter interaction is missing. Here, we show the coherent manipulation of single photons generated on-demand from a semiconductor quantum dot using a rubidium vapor quantum memory, forming a hybrid quantum network. We demonstrate the engineering of the photons' temporal wave function using four-level atoms and the creation of a new type of electromagnetic induced transparency for quantum dot photons on resonance with rubidium transitions. Given the short lifetime of our quantum dot transition the observed dynamics cannot be explained in the established steady-state picture. Our results play a pivotal role in understanding quantum light-matter interactions at short time scales. These findings demonstrate a fundamental active node to construct future large-scale hybrid quantum networks.


\end{abstract}

\pacs{Valid PACS appear here}
\maketitle

\section{Introduction}

\begin{figure*}[ht]
\includegraphics[width=\textwidth]{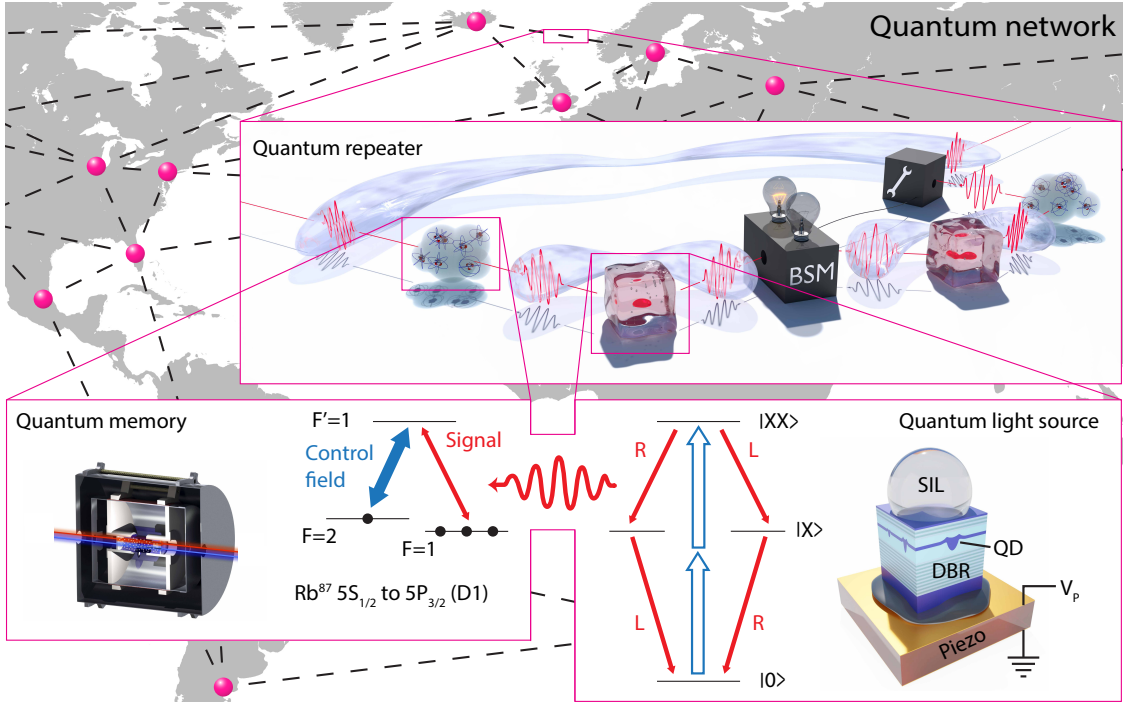}
\caption{Quantum hybrid light-matter interfaces. (Top) A schematic of a quantum hybrid light-matter interface in the context of a large quantum communication network. (Bottom right) A tunable solid-state quantum dot entangled photon pair source. The quantum dot emits entangled photon pairs with different frequencies via its biexciton-exciton cascade. The biexciton transition can be strain-tuned to emit at atomic transition wavelengths. (Bottom left) A rubidium-based room temperature quantum light-matter interface with two ground states $F=1,2$ coupled to one excited state $F'=1$ in a $\lambda$ configuration.} \label{fig:schematic}
\end{figure*}

One of the main goals of quantum communication is the transfer of quantum states over arbitrarily long distances in a quantum network. However, quantum states are very sensitive to losses in the communication channel. Several proposals exist to overcome this problem by using entanglement swapping and quantum memories. The objective of such a quantum repeater scheme is to produce a higher entanglement rate as compared to direct propagation over fiber links. The most experimentally accessible approach is based on the DLCZ protocol~\cite{Duan.Lukin.ea:2001}, that relies on quantum interference to generate entanglement~\cite{Cabrillo.Cirac.ea:1999}. However, the probabilistic nature of the scheme limits the entanglement creation rate~\cite{Moehring.Maunz.ea:2007}. An alternative scheme relies on the use of entangled photon pair sources interfaced with quantum memories, capable of receiving and storing entangled states to increase the qubit rate~\cite{Lloyd.Shahriar.ea:2001}. An improvement to the latter proposal requires the on-demand generation of polarization entangled photon pairs. A possible pathway to implement this entanglement-based quantum repeater scheme is the creation of hybrid quantum networks, in which high-performance, quantum dot based sources of on-demand entangled photon pairs~\cite{Muller.Bounouar.ea:2014,Huber.Reindl.ea:2017,Chen.Zopf.ea:2018} are interfaced with low-noise, high-fidelity atomic vapor quantum memories~\cite{Namazi.Kupchak.ea:2017,Wolters.Buser.ea:2017,Kaczmarek.Ledingham.ea:2018}. Fig.~\ref{fig:schematic} shows a schematic of the envisioned network, where solid-state quantum light sources (ice cubes) emit entangled photon pairs. One photon of each pair is stored in an atomic quantum memory (clouds) and the other photons of the pairs perform the entanglement swapping operation~\cite{Pan.Bouwmeester.ea:1998} via a Bell state measurement. 
%

%

The main challenge in interfacing these independent quantum systems is the coherent manipulation of the quantum dot emitted single photons by optically controlled atomic resonances. So far this coherent control has only been shown for \textmu{}s photons and MHz narrow atomic resonances~\cite{Matsukevich.Kuzmich:2004,Eisaman.Andre.ea:2005}. Whereas current experiments with hybrid light-matter interfaces have shown coupling of photons to 2-level atomic features~\cite{Akopian.Wang.ea:2011,Siyushev.Stein.ea:2014,Jahn.Munsch.ea:2015,Trotta.Martin-Sanchez.ea:2016,Portalupi.Widmann.ea:2016,Vural.Portalupi.ea:2018} without optically-driven coherent control.
Here, we report on the on-demand generation of photon pairs from a semiconductor quantum dot, where one photon of the pair is tuned to the rubidium D$_1$ transition, and on their coherent interaction with an electromagnetically induced transparency resonance, optically controlled in an atomic vapor. Our paper is arranged in the following sections: first we describe the components needed to build a hybrid quantum network based on a tunable GaAs quantum dot as the source, emitting single photons on-demand into the $D_1$ $F=1\rightarrow F'=1$ transition line of rubidium atoms, acting as the light-matter interface. Secondly, we characterize the optical properties of our quantum light source. Consequently, we demonstrate precise tuning of the quantum dot's emission wavelength to achieve resonance with specific rubidium transitions, determining the one-photon detuning of a single-photon driven Raman process. Lastly, we show the coherent optical manipulation of the quantum dot photons by constructing an electromagnetically-induced-transparency atomic system by placing an external diode laser tuned into two-photon resonance with the quantum dot photons. We observe spectral manipulation of the temporal wavefunction of quantum-dot photons, with lifetimes down to 134$\pm 9$\,ps, by optically manipulating the parameters of the atomic configuration.

\section{Experimental setup overview}

\begin{figure*}[ht]
\includegraphics[width=\textwidth]{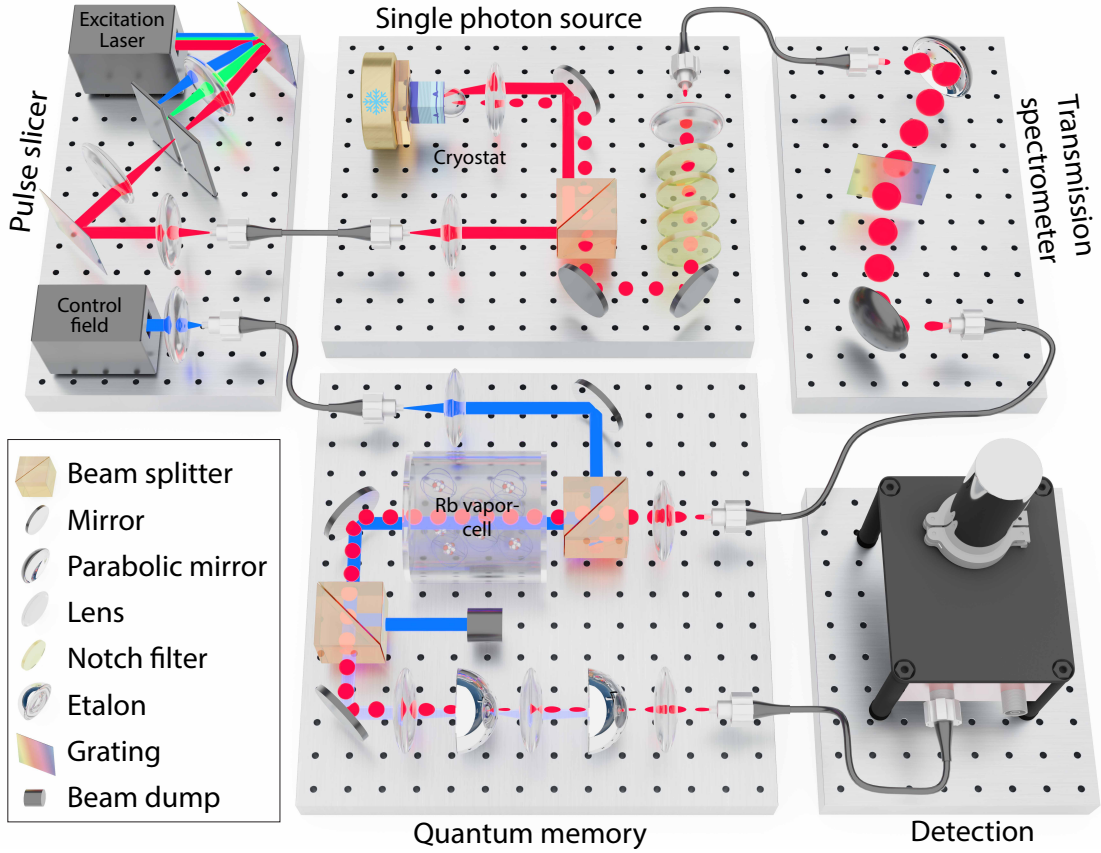}
\caption{Combined setup to interface quantum dot single photons and rubidium light-matter interfaces. From left to right: Pulsed laser system for single quantum dot excitation; Quantum dot single photon production; Spectral characterization of the quantum dot photons; Quantum memory setup including rubidium room temperature ensemble, control field and filtering stages and single photon detection after interaction. 
\label{fig:setup}}
\end{figure*}

In our experiment we realize an  active link between a quantum dot and a vapor atomic ensemble, the backbone of future hybrid quantum networks. The experiment is divided into two parts. The first part is the quantum-light source, which generates on-demand pairs of entangled photons, with one photon of the pair specifically tuned to generate quantum fields at the rubidium D$_1$ line. In the second part, we shine the temporally short single photons onto a room temperature rubidium ensemble and add a control laser field to manipulate the optical properties of the atomic medium, thereby coherently controlling the temporal wave function of the $134\pm9$\,ps short single photons. Lastly, after atomic interaction, we use a filtering system to isolate the modified photons from the control field and detect them using superconducting nanowire detectors. Figure~\ref{fig:setup} shows the different components of this hybrid quantum network, combining semiconductor, superconductor, and atomic quantum technologies. 

The ps pulses of an optical parametric oscillator laser are shaped into 10\,ps long pulses for the two-photon resonant excitation scheme~\cite{Brunner.Abstreiter.ea:1994,Stufler.Machnikowski.ea:2006}, coherently exciting the quantum dot which is located in a closed-cycle helium cryostat (5\,K sample temperature). The excitation laser is filtered using cross-polarization as well as a narrow-band notch filter. The cross-polarization is aligned to the principle axis of the quantum dot to reject one of the exciton fine-structure components. The exciton and biexciton transition are separated via a self-built transmission spectrometer with an end-to-end efficiency of 70\,\%. Afterwards the single biexciton photons are routed to a room-temperature portable quantum memory, where they are coherently transformed using engineered quantum superpositions of atomic transitions. Details on the source and the memory are given below.

\subsection{Quantum dot single photons tuned to rubidium transitions}

A desirable aspect of a single photon source to be used in a network configuration is the possibility for on-demand operation. This stringent requirement cannot be met by spontaneous parametric down conversion based sources. A recent key development along these lines has been the development of quantum light sources based on single emitters. Among them semiconductor quantum dots are at the leading edge~\cite{Senellart.Solomon.ea:2017} due to the purity of emission~\cite{Schweickert.Joens.ea:2018}, the indistinguishability of the single photons~\cite{Somaschi.Giesz.ea:2016,Ding.He.ea:2016}, the quality of the entangled quantum state~\cite{Jons.Schweickert.ea:2017,Huber.Reindl.ea:2018} as well as the possibility of ultra fast electrical operation~\cite{Boretti.Rosa.ea:2015}. 

\begin{figure*}[ht]
\includegraphics[width=1\textwidth]{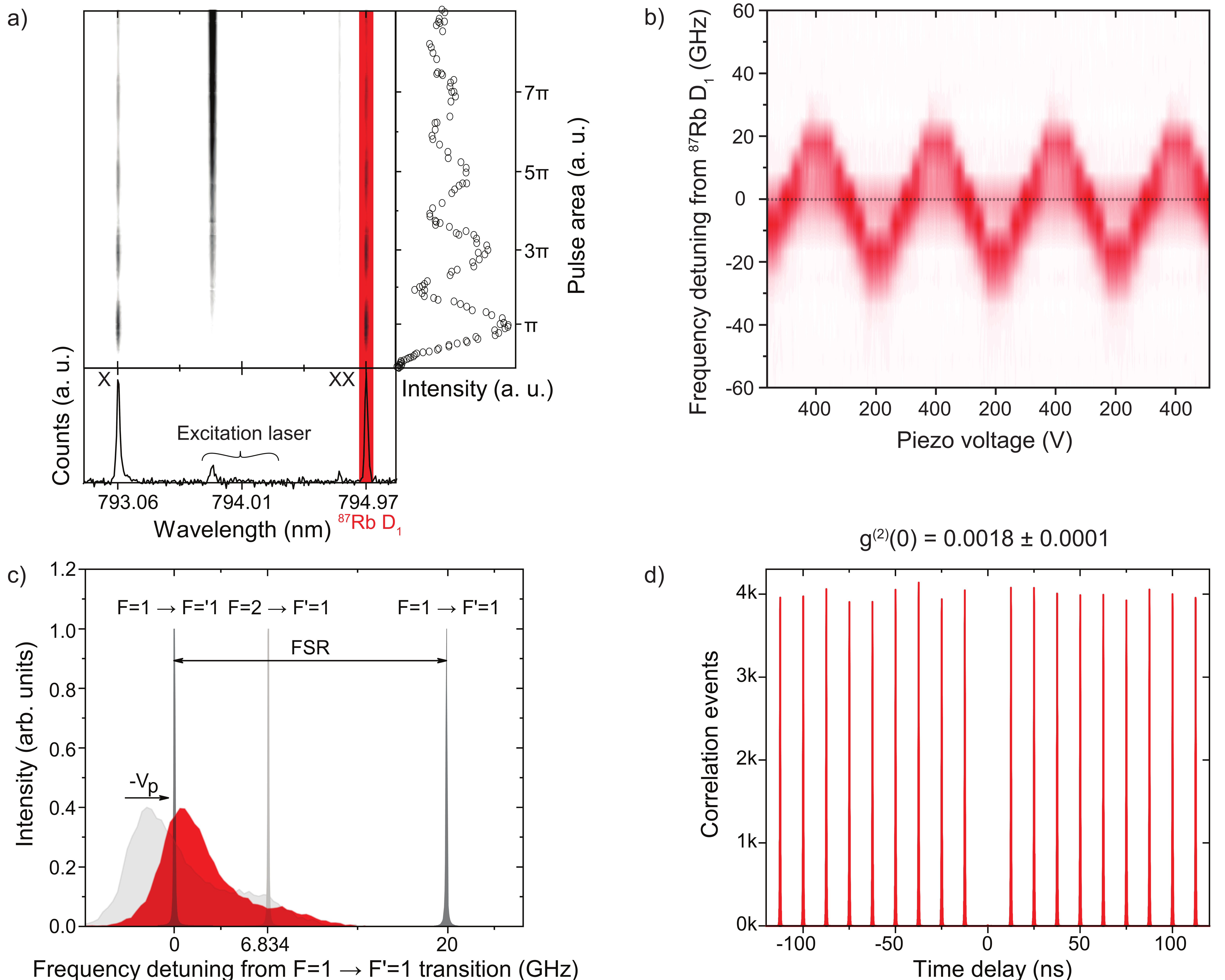}
\caption{Quantum dot light source. a) Color-coded photoluminescence spectra of the biexciton-exciton cascade under resonant two-photon excitation as a function of the excitation pulse area, revealing Rabi-oscillations. The red vertical line indicates the rubidium D$_1$ transition. b) Color-coded photoluminescence spectra of the XX transition tuned in and out of resonance of rubidium D$_1$ transition. The horizontal line is a narrow-band cw laser locked to the Rubidium D$_{1}$ line $5S_{1/2} F = 1$ $\rightarrow$ $5P_{1/2} F' = 1$ transition as a reference. c) High-resolution spectroscopy with a tunable Fabry-P\'{e}rot interferometer to precisely measure the XX transition of the quantum dot with respect to the rubidium F = 1 $\rightarrow$ F' = 1 transition.  By fine adjusting the piezo voltage (V$_P$) the XX line is brought into resonance.  d) Second-order intensity auto-correlation measurement of the XX transition under coherent $\pi$-pulse excitation, shows g$^{(2)}(0)=0.0018\pm0.0001$.\label{fig:qd}}
\end{figure*}

In our experiment we use GaAs quantum dots grown by molecular beam epitaxy with an s-shell biexciton (XX) resonance designed to be close to the D$_1$ line of rubidium at 795\,nm. Details on the sample growth can be found in the method section. We can precisely tune the XX emission wavelength by applying stress to the quantum dot structure via a piezo-electric substrate. Additional information of the device fabrication is given in the methods section. The quantum dot is excited with a two-photon resonant process in a confocal microscopy setup. Fig~\ref{fig:qd}\,a) shows color-coded photoluminescence spectra of the biexciton-exciton cascade as a function of excitation laser pulse area. The Rabi-oscillations verify the coherent control of the quantum dot system and all further measurements are performed with an excitation laser pulse area corresponding to a $\pi$-pulse. 

The most important step for tuning the quantum dot's frequency is the careful measurement of the position of the center frequency of the emission spectrum with respect to the desired rubidium absorption lines. Fig.~\ref{fig:qd}\,b) shows the tuning of the quantum dot emission by the application of a DC voltage to the piezoelectric actuator, the emission is compared to a laser locked to the rubidium D$_{1}$ line $5S_{1/2}\,F = 1$ $\rightarrow$ $5P_{1/2}\,F' = 1$ transition. Our device can be repeatably tuned in and out of resonance with a much higher precision than the resolution of the spectrometer. To verify the correct resonance tuning we thus employ a high-resolution photoluminescence spectroscopy (HRPL) setup consisting of a tunable fiber-coupled Fabry-P\'{e}rot interferometer with a resolution of 70\,MHz and free spectral range (FSR) of 20\,GHz. Fig.~\ref{fig:qd}\,c) displays four HRPL measurements. Two spectral orders of a laser locked to the F = 1 $\rightarrow$ F' = 1 and one spectral order of a laser locked to the $F = 2 \rightarrow F' = 1$ are shown as reference to determine the absolute detuning of the quantum dot XX transition to the desired Rb resonance. Two biexciton HRPL spectra are shown for different piezo tuning ($\Delta V_p = 11$\,V), highlighting our achieved tuning precision. For clarity only one spectral order of the XX emission is plotted. The tail at the right (low energy side) of the XX transition is tentatively attributed to the phonon-sideband. We extract the linewidth of the XX transition without the phonon-sideband using a Gaussian fit function, resulting in a full width at half maximum of 4.0\,GHz. The biexcitonic single photon emission of our coherently-driven quantum dot is sent to a Hanbury-Brown Twiss setup to measure the spectral correlations between subsequently emitted photons using time tagging. Fig.~\ref{fig:qd}\,d) depicts the histogram after correlating the time-tagged data to reassemble a pulsed second-order intensity autocorrelation $g^{(2)}(\Delta \tau)$-function. The distance between the side peaks correspond to the laser repetition rate of ~80\,MHz. As a result a g$^{(2)}$($\tau$) correlation function is measured with a final g$^{(2)}(0)=0.0018\pm0.0001$, showing almost perfect single photon emission. Several quantum dots have been similarly characterized, showing comparable optical properties: less than a factor of 2 away from the Fourier-limit, XX lifetimes around 130\,ps, wavelength tunability of 0.5\,nm, extremely low multi-photon emission, and small fine-structure splitting. 

\subsection{Room temperature quantum light-matter interface}
In our vision of a hybrid quantum networks connecting distant locations, portable and robust quantum light-matter interfaces serving as buffers and memory are of paramount importance. In the next part of the experiment we use a first-prototype of a fully-portable plug-and-play room temperature quantum memory. This prototype has the same features of the designs used in previous experiments~\cite{Namazi.Kupchak.ea:2017}, but now being fully independent of laboratory infrastructure, only requiring probe photons and control fields as inputs. 

We collect the quantum dot photons tuned to the $5S_{1/2}\,F = 1$ $\rightarrow$ $5P_{1/2}\,F' = 1$ transition with an polarization maintaining fiber serving as a communication quantum channel reaching the quantum memory. The quantum memory is based upon a warm $^{87}$Rb vapor and controlled using electromagnetically induced transparency (EIT). We employed an external-cavity diode laser, locked to the $5S_{1/2}\,F = 2$ $\rightarrow$ $5P_{1/2}\,F' = 1$ rubidium transition using a saturated absorption setup and the Pound-Drever-Hall technique, as the control field. The control beam coherently prepares a volume within a $^{87}$Rb vapor cell at $60^{\circ}\,$C, containing Kr buffer gas to serve as the atomic medium to transform the short single photons. 

The quantum memory also possesses a miniaturized version of the filtering system as compared to our previous implementation
with independent temperature controllers. Polarization elements supply 42\,dB of control field attenuation ($\approx$\,90\,\% probe transmission) while two temperature-controlled etalon resonators (linewidths of 43 MHz) provide additional 102\,dB of control field extinction. The total probe field transmission is maximally 15\,\%, exhibiting an effective, control/probe suppression ratio of 136\,dB. After filtering, another fiber brings the resulting transmitted quantum dot photons to the detection system. The photons are detected using superconducting nanowire single photon detectors with ultra low noise ($(0.017\pm0.001)$\,\si{\per\second}) and high quantum efficiency of \SI{64}{\percent}.

\section{Quantum model of fast light-matter interaction}
The crucial ingredient to convert the hybrid quantum interconnects from ``passive" absorptive systems to ``active" light-controlled devices is to understand the interaction between temporal short single-photons and coherently prepared multi-level atomic transitions. This will allow to tailor novel light-matter interaction mechanisms to achieve desired functionalities specific to the hybrid architecture. In this section we present a numerical model describing the interaction of a fast single photon excitation with a thermal ensemble of four-level rubidium atoms.

We model the quantum dot single photon emission in the following manner. First we assume a fundamental quantum dot emission following a Lorentzian frequency profile:
\begin{eqnarray}\label{eq_qDotLorentzian}
	|\boldsymbol{E}(\omega)|^2 = \frac{1}{\pi} \frac{\gamma}{(\omega-{\omega}_p)^2+\gamma^2}
\end{eqnarray}
where $\gamma=\frac{\Gamma}{2}=\frac{1}{2\tau}$ is the half decay rate of the quantum dot energy levels.

The time profile of the single photons has thus the form:
\begin{eqnarray}\label{eq_probeTemporal}
	\boldsymbol{E}(z,t) = i\sqrt{\frac{\hbar\omega_p}{2V\epsilon_0}} \,a\,(z,t) e^{-i \omega_p t} \theta(t) \,\boldsymbol{\epsilon} +c.c.
\end{eqnarray}
where $\boldsymbol{\epsilon}$ is the unitary polarization vector. The used parameter is $\gamma = 3.73$ GHz, corresponding to decay time of $\tau = 134$\,ps. At the boundary where the quantum dot photons enter the atomic ensemble, the field operator has the simple form: $a(z=0,t) := e^{-\gamma t}$.

The Rotating-Wave-Approximation Hamiltonian of four level atoms interacting with the probe and control fields is:
\begin{equation}\label{eq_hamiltonian}
\begin{aligned}
H/\hbar &= (\Delta_c-\Delta_p) \sigma_{22} - \Delta_{p}  \sigma_{33} \\
&+ (\omega_{43}-\Delta_p)\sigma_{44} + \left(g_{p1} \sigma_{31} + g_{p2}\sigma_{41}\right)  a \\
&+  \left(g_{c1} \sigma_{32} + g_{c2}\sigma_{42}\right) b + c.c. 
\end{aligned}
\end{equation}
where we have defined $\omega_{11}:=0$, and used the following definitions: $\Delta_p:=\omega_p-\omega_{31}$, $\Delta_c:=\omega_c-\omega_{32}$. 

The complete temporal wave function of the propagating photons can be obtained by having the atomic states evolving together with the photonic states according to a Maxwell-Bloch equation:
\begin{equation}\label{eq_propagation}
\begin{aligned}
\left(\frac{\partial}{\partial t} + c \frac{\partial}{\partial z}\right) {a}(z, t) &= i N \left( g_{p1}\rho_{31} (z, t)\right.\\
&\left.+g_{p2}\rho_{41} (z, t)\right).
\end{aligned}	
\end{equation}
where  $N=lAn$ is the total number of atoms in the cylindrical volume prepared by the control field, and the density of atoms is calculated through
$n=\frac{133.323}{k_B T}\times10^{-94.04826-\frac{1961.258}{T}-0.03771687T+42.57526\log_{10}T}$. The used length of cell is $l=5$ cm, and the beam cross section is estimated to be $A=3.14$ mm${^{2}}$.

Using as starting point the temporal wave-function of the photon emitted by the quantum dot, we numerically evaluate the quantum dynamics of the pulse in the medium by solving the aforementioned equations for specific one-photon detunings. The final response of the systems should also include broadening effects due to thermal motion of the rubidium atoms. This one-photon detuning frequency distribution is modeled using a Gaussian function, centered around the main rubidium resonance. 
\begin{eqnarray}\label{eq_qDotFreq}
	P(\omega_p) := \frac{1}{\sqrt{2\pi\sigma_p^2}}\,e^{-\frac{(\omega_p-\omega_0)^2}{2\sigma_p^2}}
\end{eqnarray}

\section{Experimental results}
\subsection{Interaction of quantum dot photons with rubidium atoms.}
In our first set of experiments we shined the rubidium-tuned quantum dots photons onto the rubidium quantum memory and measured the changes in the temporal wave function of the single photons due to the interaction with the atomic ensembles for different temperatures of the rubidium cell. We measure the resultant temporal wave functions by creating time-binned histograms of the transmitted and detected photons after atomic interaction at resonance. 
 
\begin{figure*}[htb]
	\includegraphics[width = 1\textwidth]{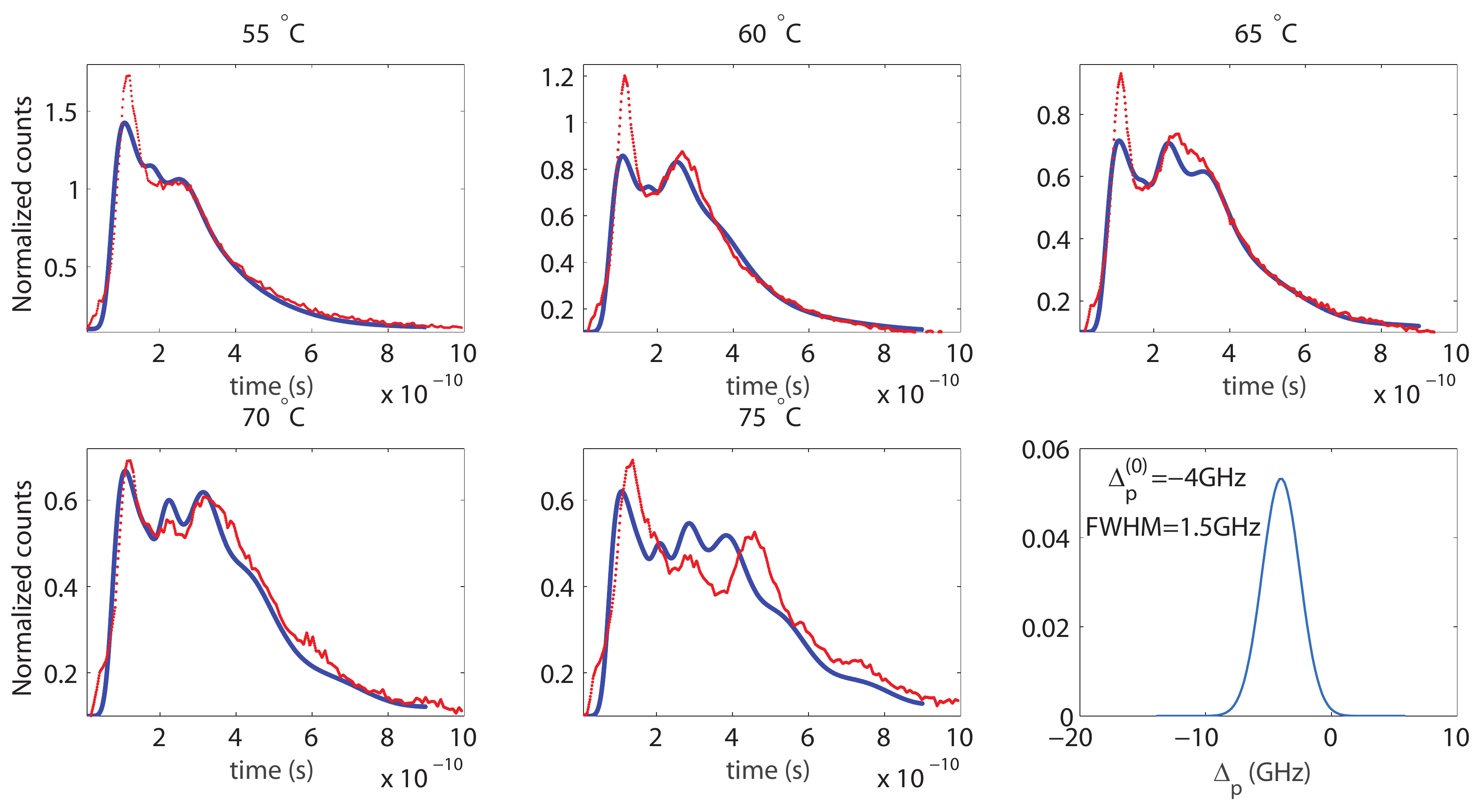}
	\caption{On-resonant light-matter interaction of 134\,ps long single photons with an ensemble of rubidium atoms at different optical densities. Figures show the results of applying the light-matter interaction model (blue line) to the quantum dot single-photon temporal wave function. Red dots are a histogram of time-correlated single-photon counting measurement. (bottom right) Frequency distribution used to evaluate the effect of Doppler broadening in the light-matter interaction model. The numerical parameters are kept constant and only the atomic density is changed. We have used the following parameters: beam waist is $w=2$\,mm, atomic decay rates $\Gamma_{31}=\Gamma_{32}=2\pi\times 1.5$ MHz. The QD photon frequency centered around -4\,GHz blue detuning from $^{87}$Rb D$_1$ $F=1\rightarrow F'=1$ transition, and a Doppler width of the atoms of 1.5\,GHz. More details on the simulations are given in the method section.}
	\label{fig:tempseries}
\end{figure*}

The D$_1$ line in rubidium 87 atoms contains two ground states (F=1 and F=2) and two many-folds of transitions to each excited states (F'=1 and F'=2). With a Doppler width of 500\,MHz and a pressure broadening of 300\,MHz, both broadened many-fold resonances (separated by 6.834\,GHz) have a full width half maximum resonance of 800\,MHz. This compared to the $\approx$ 4.0\,GHz Gaussian FWHM of the quantum dots photons produces significant quantum light-matter dynamics. These measurements are shown as red dots in Fig.~\ref{fig:tempseries}, for different temperatures.

We observe fundamentally different dynamics in this resonant excitation case as compared to the off-resonant interaction reported in many previous implementations~\cite{Akopian.Wang.ea:2011,Trotta.Martin-Sanchez.ea:2016,Vural.Portalupi.ea:2018}. In previous observations of hybrid light-matter interaction, the output of the system can be modeled as a ``leakage" and a ``slow down part", if the center frequency of the dot emission is placed in between the F = 1 and F = 2 ground state absorption many-folds. In the resonant case explored here, we observed the creation of several time-dependent wave packets associated to fast oscillations in the atomic excitation. In Figure~\ref{fig:tempseries}, we see the delay between these wave packets increasing with atomic density (red dots in Fig.~\ref{fig:tempseries}). This rich quantum dynamics is in full agreement with the predictions of our numerical model (blue). We emphasize that in the computational trends presented on Figure~\ref{fig:tempseries}, all simulation parameters remain constant, while we only vary the atomic density through temperature changes. We consider this to be a good benchmarking of the dynamics predicted by the model with respect to the experimental data.

\subsection{Fast electromagnetically induced transparency with quantum dot single photons}
The basis to achieve quantum coherent control in an atomic ensemble is the creation of two-photon Raman resonances. Single photon/control field two-photon resonances are at the heart of the coherent photon manipulation in atomic techniques such as electromagnetically induced transparency and off-resonant Raman interaction. In our case, we have constructed a system in which very short photons interact at two-photon resonance with four-level atoms. We achieve this two-photon resonance by adding a control field resonant to the $F=2 \rightarrow F'=1$ transition to the quantum dot single photons already resonant to the $F=1 \rightarrow F'=1$ atomic transition. 

\begin{figure*}[ht]
	\includegraphics[width = 1\textwidth]{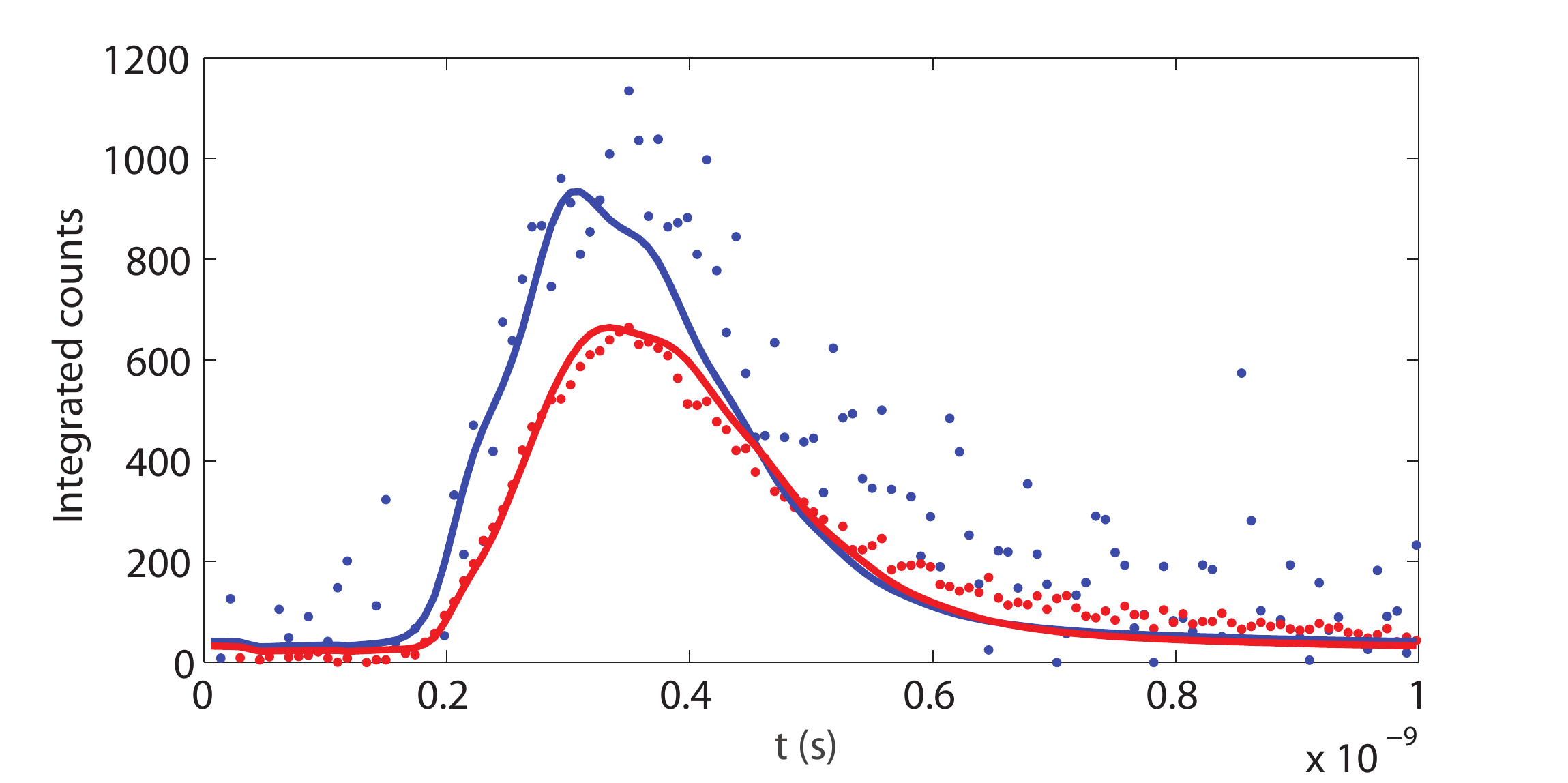}
	\caption{Fast electromagnetically-induced transparency. We show histograms of arrival time for photons that are transmitted through the atomic system and the etalon filtering system. The red dots indicate the field that interacted with the atoms but no control field. The blue dots indicate quantum dot photons that see a two-photon EIT resonance, demonstrated by the optical control of the transparency. Numerical simulations used as starting parameters the simulation presented in Fig. 4 for the data at $T=60^\circ{C}$ plus the transmission through the filtering system (red solid line). The FSR of the employed etalon is 13.3 GHz, with line width 43 MHz. The EIT simulation (blue solid line) is obtained with the same parameters and an extra control field $\Omega_{c}$, equivalent to $3\times 10^5$ V/m.}
\label{fig:eit} 
\end{figure*}

In the experiment the quantum dot photons and the control field have orthogonal polarizations and are overlapped using a Glan-laser polarizer. Due to the presence of the control field and the photons, the atoms are prepared in a superposition of the ground states $F=1$ and $F=2$. This situation resembles the usual static EIT case. However, as a consequence of the short duration of the pulses, the quantum dynamics of the time-dependent interaction differs from that of EIT driven by long photonic pulses. We call this new phenomenon fast-EIT. 

After atomic interaction, the fields are separated again using polarization elements and two etalon frequency filters. This allows the measurement of exclusively the quantum dot photons that interacted in the fast-EIT configuration. In Fig.~\ref{fig:eit} we show the fast electromagnetically induced transparency effect for quantum dot photons (blue dots) and its comparison to experiments without the control field (red dots). As it can be seen, the presence of a control field increases the transmission, compared to the case with resonant transmission without control field, demonstrating the laser-driven coherent control of the fast-EIT quantum dynamics. Each data set was integrated for 10\,h in a time correlated single-photon counting measurement. The measurement with control field on (blue dots) had a signal rate of $\approx$ 5\,cts./s and $\approx$ 2000\,cts./s uncorrelated cw background photons generated by the control field. The cw background was subtracted from the data for comparison. The measurement without control field (red dots) had a signal rate of $\approx$ 3\,cts./s and 0.02\,cts./s dark count contribution from the superconducting single-photon detector.

In order to extend our simulation to the fast-EIT case, we have to add two considerations. Firstly, we must add the frequency-filtering effect due to our etalons system on the temporal wave-function of the photons after light-matter interaction. We do this by using the etalons parameters (R=99.9,\,$\nu_{FSR}=13.3$ GHz, $\Gamma_E = 43$ MHz). We take as input the temporal wave-function of the photons, as calculated in  the fit to the data in Fig. 4 (blue line for the 60$^{\circ}$C measurement). We Fourier transform this time-domain wave-function to obtain the frequency domain information using: $a(\nu) =  \mathcal{F}_t\left[ a(t) \right](\nu)$. Then we use the complex amplitude transfer function for a single FSR \cite{Yu:01}, defined as $tr(\nu)=e^{-i\pi \nu/\nu_{FSR}} \frac{1-R}{1-R\,e^{-2i\pi \nu/\nu_{FSR}}}$. Finally, the filtered field is obtained using 
\begin{eqnarray}
a(t) = \mathcal{F}_\nu^{-1}\left[ a(\nu)\, tr(\nu)\right](t)
\end{eqnarray}
and the intensity is calculated as $I(t) = |a(t)|^2$. Figure 5 (red line) shows the results of this procedure. This full simulation completely resembles the experimental data. 

Lastly, we repeat the procedure used to obtain the simulations presented in Fig.~\ref{fig:tempseries}, but now with a control field switched on, following the dynamics outlined by the aforementioned four-level atoms Hamiltonian. The outcome of this simulation also undergoes the filtering procedure described above and we obtain a final prediction (blue line in Fig.~\ref{fig:eit}) on the modification of the temporal wave function of the photons, which is also in excellent agreement with the experimental data and shows an electromagnetically induced increase in transmission by 27.76\,\%. The difference between simulation and experiment might stem from instability during the long integration time of the measurement.

\section{Discussion}
We have found a novel regime of coherent dynamics in which the atomic manipulation of short quantum dot photons is possible. We have modeled the new phenomenon quantum mechanically and demonstrated experimentally the predictions of our theory. This novel form of optical quantum coherent control will be the basis to design new hybrid quantum devices based upon the optical manipulation of short photons produced in condensed-matter systems, in optically-controlled atomic ensembles. 

In the context of the creation of large quantum hybrid networks, this novel approach to manipulate short photons is of great interest. It leverages two of the most versatile quantum devices: i) the best available on-demand photon pair source, with high photon pair generation efficiency and fine tuning to match the frequency of the biexciton quantum dot photons with the $D_1$ $F=1\rightarrow F'=1$ rubidium transition, and ii) portable room temperature rubidium quantum light-matter interfaces addressable by controlling laser fields. This hybrid link between two independent ``quantum hardware'' elements could be the basis for important applications in quantum technology. 
 
These new hybrid interconnects still require technical upgrades before they are able to fulfill their complete potential. Due to the still existing bandwidth mismatch, only a fraction from the quantum dot photons coherently interacted with the four-level atoms, which remains one of the main challenges for the efficient manipulation and storage of single-photons. A possible route to overcome this final step from both ends is (i) the quantum dots need to emit Fourier-limited photons under two-photon resonant excitation, which might be possible by further improving the material quality and cooling the sample further and (ii) the bandwidth of the quantum memory has to be increased, using large-bandwidth Raman off-resonant memory schemes instead of on-resonance EIT-like techniques. 

Our experiment marks an important step forward to realize hybrid communication and processing quantum networks. Due to its simplicity, it now possible to create networks of multiple room temperature quantum memories. Current improvements in cryogenic technology also allow to field many scalable polarization entangled sources remotely. We envision that in the short term, all the required functional building blocks to demonstrate all-photonic hybrid quantum networks will reach their technological maturity.


\section{Methods}
\label{method}

\subsection{Quantum dot sample} 
\label{method:sample}
The quantum dot sample was grown by molecular beam epitaxy at the JKU in Linz. First a bottom distributed Bragg reflector made of 9 pairs of $\lambda$/4-thick Al$_{0.95}$Ga$_{0.05}$As (\SI{68.9}{\nano\meter}) and Al$_{0.2}$Ga$_{0.8}$As layers is deposited as a mirror for a $\lambda$-cavity. The quantum dot layer is located at the center of the $\lambda$-cavity made of a $\lambda$/2-thick (\SI{123}{\nano\meter}) layer of Al$_{0.4}$Ga$_{0.6}$As sandwiched between two $\lambda$/4-thick (\SI{59.8}{\nano\meter}) Al$_{0.2}$Ga$_{0.8}$As layers. The quantum dot layer is fabricated by Al-droplet etching~\cite{Heyn.Stemmann.ea:2009,Huo.Witek.ea:2014} on the Al$_{0.4}$Ga$_{0.6}$As layer followed by deposition of \SI{2}{\nano\meter} GaAs. The top mirror for the cavity is made of two pairs of the same material combination as the bottom distributed Bragg reflector. To protect the structure a 4 nm-thick GaAs protective layer covers the final sample. The QD design emission wavelength is centered around $\sim$\,\SI{790}{\nano\meter} and a gradient in the mode position (Q factor of about 50) is generated by stopping the substrate rotation during the top Al$_{0.2}$Ga$_{0.8}$As cavity-layer deposition. The cavity design together with solid immersion lens enhances the collection efficiency by $\sim$\,30 times compared to an unstructured sample.

\subsection{Piezo-actuator integration} 
\label{method:piezo}
To achieve stable and precise energy tuning of our quantum dot emission into resonance with the rubidium $D_1$ $F=1\rightarrow F'=1$  transition we integrate the GaAs quantum dot structure on a gold coated PMN-PT piezoelectric actuator (TRS technologies, thickness of $200\,\si{\micro\meter}$, $\langle 001 \rangle$ orientation). The quantum dot sample is mechanically thinned using diamond-based abrasive films and transferred on the piezo-electric substrate  with a bendable soft tip in pick and place method, taking advantage of electrostatic forces. We use cryogenic epoxy (Stycast, two component resin) resulting in a rigid connection between the sample and the piezo-actuator with good thermal contact. More details on the quantum dot sample transfer on the piezo-electric substrate can be found in Ref.~\onlinecite{Zeuner.Paul.ea:2018}.

\subsection{Numerical methods}
To obtain the information of the input pulse serving as the starting parameters of the simulation, we select a segment of 1.5 ns from the raw single photon count histograms, which covers the quantum dot photons temporal envelope. Noise sources are removed by using a Lorentzian fitting of the photon pulse.
\begin{eqnarray}
f_1 = \frac{q_1 q_3}{(t-q_2)^2+q_1^2} + q_4
\end{eqnarray}
with the following parameters:
\begin{center}
\small
    \begin{tabular}{ | c | c | c | c | c | c |}
    \hline
    T ($^\circ{C}$) & 55 & 60 & 65 &  70 & 75 \\ \hline
	$q_1$ (10$^{-10}$) & 0.4122  &  0.4070  &  0.4353  &  0.4365  &  0.4590\\ \hline
    $q_2$ (10$^{-7}$) & 0.1490  &  0.1490  &  0.1490  &  0.1490 &   0.1491 \\ \hline
    $q_3$ (10$^{-6}$) & 0.1449 &   0.1874  &  0.2385  &  0.3282  &  0.4643 \\ \hline
    $q_4$ & 0.9900  &  0.9900  &  0.9900  &  0.9900   & 1.0233
    \\ \hline
    \end{tabular}
\end{center}
Next we identify the decay time of quantum dot photons, by performing a logarithmic fitting of the processed data described above. We find a value of $\tau=134$ ps.

The main parameters in actual simulation are listed on Table \ref{tb_parameters}.
The quantum dot detuning frequency range was first discretized from -13.9 GHz detuning to +5.9 GHz into 100 individual values, and then the response time profile for each detuning frequency was calculated.
After collecting all the information of individual detuning responses, we then build a Gaussian distribution model to weight over possible detuning, following Eq.\ref{eq_qDotFreq}.

The calculation of one detuning frequency can be described as follows. For one detuning frequency $\Delta_p=\nu_p-\nu_{31}$ of the quantum dot  photon at a specific temperature T, we run a complete Hamiltonian simulation.

We add the effect of Doppler broadening in the rest frame of the atoms according to a Boltzmann distribution. We discretized the atomic velocity in the range $\pm\sqrt{8 \ln2 k_B T/m}$ into 14 values $v_n, n=-7,...,7$. And for each $v_n$ the program is run separately, under the condition that $\Delta_p^{Doppler}=\Delta_p+\nu_pv_n/c$, $\Delta_c^{Doppler}=\Delta_c+\nu_{32}v_n/c$; then all 14 Doppler broadened responses are weighted with Maxwell's distribution function of 1D velocity $\frac{\exp{-mv_n^2/(2k_BT))}}{\sum_{n=-7}^7 \exp{-mv_n^2/(2k_BT))}}$. 

This leaves us the core unit of simulation, which will explain how we process the calculation for one detuning frequency at one temperature with one specific Doppler broadening.
This part consists of iteratively solving Maxwell-Bloch equation \eqref{eq_propagation} that describe propagation of qDot photon field dynamics, and master equation 
\begin{equation}\label{eq_masterEquation}
\begin{aligned}
\dot{\rho} &= \frac{1}{i\hbar} \left[ H,\rho \right] \\
&+ \sum_{n=3,4}\sum_{m=1,2} \Gamma_{nm} \left( 2\sigma_{mn} \rho \sigma_{nm}  - \sigma_{nn}\rho - \rho \sigma_{nn} \right)
\end{aligned}
\end{equation}
where $\Gamma_{nm}=2\pi\times1.5$ MHz, which describes evolution of atomic states.
In the actual simulation, Eq.\eqref{eq_propagation} was discretized as
\begin{equation}\label{eq_discretized_propagation}
\begin{aligned}
a(n_t,n_z)&=\frac{1}{1/dt+c/dz}\Big(\frac{c}{dz}a(n_t,n_{z}-1) \\
&+ \frac{a(n_{t}-1,n_z)}{dt }\\
&+ ilAn(g_{P1}\rho_{31}(n_{t}-1,n_{z}-1) \\
&+ g_{P2}\rho_{41}(n_{t}-1,n_z-1))\Big) 
\end{aligned}
\end{equation}
where $n_t=1,...,747$ are indices of discretized time of 1 ns, meaning a time difference $dt = 1.34$ ps. And we use $dz = c\, dt$ to guarantee the stability of numerical solution.
Moreover, master equation \eqref{eq_masterEquation} is vectorized and discretized to a super density matrix master equation:
\begin{eqnarray}
\dot{\Theta}_{n_t,n_z} = \mathcal{H}_{n_t,n_z} \Theta_{n_t-1,n_z}
\end{eqnarray}
where $\Theta = [\rho_{11},\allowbreak \rho_{21},\allowbreak \rho_{31},\allowbreak \rho_{41},\allowbreak \rho_{12},\allowbreak \rho_{22},\allowbreak \rho_{32},\allowbreak \rho_{42},\allowbreak \rho_{13},\allowbreak \rho_{23},\allowbreak \rho_{33},\allowbreak \rho_{43}, \rho_{14},\allowbreak \rho_{24},\allowbreak \rho_{34},\allowbreak \rho_{44}]^T$, and $\mathcal{H}$ is superoperator form of Hamiltonian and Lindbladian in Eq.\eqref{eq_masterEquation}.

\begin{table*}[ht]
    \begin{tabular}{ | c | m{7cm} | c |}
    \hline
    Parameters & Note & Value used in simulation \\ \hline
    $\hbar$ & Planck's constant & 1.0546$\times 10^{-34}$ Js \\ \hline
    $c$ & speed of light in vacuum &299792458 ms$^{-1}$ \\ \hline
    $k_B$ & Boltzmann's constant &$1.380 \times 10^{-23} m^2 kg s^{-2} K^{-1}$ \\ \hline
	$m$ & Rb 87 atomic mass & $1.443194628\times 10^{-25}$\\ \hline
    $\epsilon_0$ & vacuum permittivity &$8.85\times 10^{-12} Fm^{-1}$ \\ \hline
    $l$  & cell length & 5 cm \\ \hline
    $w$ & beam waist & 2 mm \\ \hline
    $A$ & beam cross section & $\pi(w/2)^2$ \\ \hline
    $\omega_{31}$ & F=1$\rightarrow$ F'=1 & $2\pi\times377.1112248$ THz \\ \hline
    $\omega_{32}$ & F=2$\rightarrow$ F'=1 & $2\pi\times377.1043901$ THz \\ \hline
    $\omega_{43}$ & between F'=2 and F'=1 & $2\pi\times816.65630$ MHz \\ \hline
    $d$ & dipole moment & $2.537\times 10^{-29}$ Cm\\ \hline
    $d_{11}$ & dipole moment F=1$\rightarrow$ F'=1 & $d/\sqrt{6}$\\ \hline
    $d_{12}$ & dipole moment F=1$\rightarrow$ F'=2 & $d/\sqrt{1.2}$\\ \hline
    $d_{21}$ & dipole moment F=2$\rightarrow$ F'=1 & $d/\sqrt{2}$\\ \hline
    $d_{22}$ & dipole moment F=2$\rightarrow$ F'=2 & $d/\sqrt{2}$\\ \hline
    $g_{P1}$ & coupling coefficient between atomic transition F=1$\rightarrow$ F'=1 and qDot photon & $d_{11}\sqrt{\frac{\omega_{31}}{2lA\epsilon_0\hbar}}$\\ \hline
    $g_{P2}$ & coupling coefficient between atomic transition F=1$\rightarrow$ F'=2 and qDot photon & $d_{12}\sqrt{\frac{\omega_{31}}{2lA\epsilon_0\hbar}}$\\ \hline
    $g_{C1}$ & coupling coefficient between atomic transition F=2$\rightarrow$ F'=1 and control field & $d_{21}\sqrt{\frac{\omega_{32}}{2lA\epsilon_0\hbar}}$\\ \hline
    $g_{C2}$ & coupling coefficient between atomic transition F=2$\rightarrow$ F'=2 and control field & $d_{22}\sqrt{\frac{\omega_{32}}{2lA\epsilon_0\hbar}}$\\ \hline
    \end{tabular}
    \caption{Physical constants and equations used in simulation.}
    \label{tb_parameters}
\end{table*}

\clearpage

\section{Acknowledgements}
K.D.J. acknowledges funding from the MARIE SK\L ODOWSKA-CURIE  Individual  Fellowship under REA grant agreement No. 661416 (SiPhoN). This work was financially supported by the European Research Council (ERC) under the European Union’s Horizon 2020 research and innovation programme (SPQRel, Grant Agreement 679183) and the European Union Seventh
Framework Program 209 (FP7/2007-2013) under Grant Agreement No. 601126 210 (HANAS). The JKU group acknowledges the Austrian Science Fund (FWF): P29603 and V. Volobuev, Y. Huo, P. Atkinson, G. Weihs and B. Pressl for fruitful discussions. The KTH group acknowledges the continuous support by the companies APE Angewandte Physik und Elektronik GmbH on their picoEmerald system and Single Quantum BV on their SSPDs. The Stony Brook group acknowledge funding from the National Science Foundation, grants number PHY-1404398 and PHY-1707919 and the Simons Foundation, grant number SBF241180.

\section{Author contribution}
K.D.J., E.F., and V.Z. conceived and designed the experiment. L.S. and K.D.J build the quantum dot source setup with the help of K.D.Z. and T.L.. M.N. and E.F. designed the quantum memory and L.S., K.D.J., E.F. set it up and optimized it with the help of L.S.M.. L.S. and K.D.J carried out the experiments and analyzed the data together with K.D.Z. and E.F.. G.C. and E.F. wrote the theory and performed the simulations. S.F.C.S., H.H., and A.R. grew the quantum dot sample. M.R. and R.T. characterized the quantum dot sample and T.L., L.S., and K.D.J. fabricated the piezo-tunable device. V.Z. and E.F. supervised the project. K.D.J. and E.F. wrote the manuscript with inputs from all authors.
\bibliographystyle{unsrtnat}
\bibliography{quantumnetwork}
%





\end{document}